# Effect of Climate and Geography on Worldwide Fine Resolution Economic Activity

Short title: Effect of geography on fine resolution economy

Author: Alberto Troccoli[a,1]

[a] World Energy and Meteorology Council, The Enterprise Centre, University of East Anglia, Norwich, NR4 7TJ, UK

[1] To whom correspondence should be addressed. Email: alberto.troccoli@wemcouncil.org

# Abstract

Geography, including climatic factors, have long been considered potentially important elements in shaping socio-economic activities, alongside other determinants, such as institutions. Here we demonstrate that geography and climate satisfactorily explain worldwide economic activity as measured by the per capita Gross Cell Product (GCP-PC] at a fine geographical resolution, typically much higher than country average. A 1° by 1° GCP-PC dataset has been key for establishing and testing a direct relationship between 'local' geography/climate and GCP-PC. Not only have we tested the geography/climate hypothesis using many possible explanatory variables, importantly we have also predicted and reconstructed GCP-PC worldwide by retaining the most significant predictors. While this study confirms that latitude is the most important predictor for GCP-PC when taken in isolation, the accuracy of the GCP-PC prediction is greatly improved when other factors mainly related to variations in climatic variables, such as the variability in air pressure, rather than average climatic conditions as typically used, are considered. Implications of these findings include an improved understanding of why economically better-off societies are geographically placed where they are.

# Introduction

The relevance of meteorology (comprising weather and climate) and geography variables on the socio-economic activities is a fascinating subject and one that has been touched on intermittently for centuries. Notable mentions are Montesquieu's treatise 'The Spirit of Laws' of 1750 in which, as reported by Dell et al. (2012), he argued that an 'excess of heat' made men 'slothful and dispirited'. More than a century and a half later, Huntington (1915) analysed in more detail how key weather variables such as temperature and humidity influence labour productivity. He found that not only did temperature have an effect on labour productivity, also other meteorological factors such as relative humidity, wind speed, and storms (or pressure changes) were important too. He also considered large scale patterns and speculated that changes in the global civilisation might be linked to centennial variations in climate. It is this macro-scale approach, rather than Huntington's more specific but motivational labour productivity work, that is the focus of this work.

There is no doubt that economic development is a highly complex phenomenon, one which inextricably links physical and social factors. It is nonetheless possible, and highly interesting, to investigate how individual determinants contribute to it (e.g. Bloom and Sachs 1998, Rodrik et al. 2004). While full consideration has been given to representing a wide range of determinants, this work focuses on the role of geography and climate for reasons explained below.

Only sporadic studies followed the work of Huntington, until the last few decades when Diamond (1997), Bloom and Sachs (1998), Nordhaus (2006), Dell et al. (2014) re-elaborated and advanced our understanding about the importance of geography and climate in economic-growth studies. Possibly the main reason for the sporadic studies in the twentieth century is that these *became associated with racism because it intimated that people in the tropics were less productive than those in the temperate zone* (Glantz 2003). However, the use of improved data and methodologies in the last few decades have led to much more solid research results and therefore rendered that criticism seemingly anachronistic. Investigations that have taken into account geography/climate to explain socio-economic activity normally use an invariant geography variable, latitude, and just a few climatic variables, mainly mean temperature and precipitation, as explanatory factors (Rodrik et al. 2004, Dell et al. 2012, Burke et al. 2015). The corollary of this is that latitude, mean temperature and precipitation (individually or in combination) essentially have become synonym of geography/climate in a large component of the current literature in this area. This has crucial implications because, albeit important, mean temperature and precipitation are just two of the variables regulating weather and climate. Therefore there is a risk to dismiss the geography/climate argument just because it is not always possible to explain economic activity by just using these variables. That is why it is critical to consider meteorology, together with geography, in a more holistic way, as we do in this work.

One of the limited examples where more geography/climate variables are considered is for the case of African economic activity (Nordhaus and Chen 2009). By using linear and squared terms in mean precipitation, mean temperature, elevation, and the distance from coastline, lakes, and rivers, they concluded that these explain a substantial proportion of the economic output for Africa. In Mitton (2016) determinants of economic development covering 1867 subnational regions from 101 countries, focusing on *within*-country effects of geography and institutions using 25 geography and mean climate variables were investigated. It was concluded that while institutions have a significant positive effect on income among subnational regions with greater autonomy, he found that, *simply put, geography matters*.

The question this work addresses is: "*What is the role of climate/geography in worldwide fine resolution economic activity and specifically which climatic/geographical variable is most relevant for economic activity?*" To achieve this, we considerably extend the number, and crucially the type, of geography/climate variables used by Nordhaus and Chen (2009) and Mitton (2016). More specifically, in addition to the mean of variables such as temperature, precipitation, air pressure, relative humidity, dew point temperature, wind speed, solar radiation and sunshine duration, we consider their variations in time. The idea behind the use of variations in climate variables is that we can for instance mimic the effect of decreasing air pressure, which is akin to an incoming weather perturbation, such as a storm. This approach allows us to assess the effect of wider meteorological drivers on the worldwide economic activity, as measured by the per capita Gross Cell Product (GCP). The conceptual basis of GCP is the same as that of gross domestic product (GDP) and gross regional product as developed in the national income and product accounts of major countries, except that the geographic unit is the latitude-longitude grid cell (Nordhaus and Chen 2009).

This work contributes to the debate about the role geography and climate have in shaping communities and economies. Such a debate has been driven by two fundamental determinants: institutions versus geography/climate (some authors include also a third determinant, international trade or market integration, e.g. Sachs and Warner [1995], and Frankel and Romer [1999], which is however itself dependent on institutions, and geography). On the one hand, the popular book by Acemoglu and Robinson (2012), and before it North (1990), Rodrik et al. (2004), Bosker and Garretsen (2009), have been arguing about the criticality of the role of institutions in driving productivity. Rodrik et al. (2004) attempted a quantification of the respective roles of institutions, geography/climate (simulated by latitude only), but also integration, and concluded that *once institutions are controlled for, integration has no direct effect on incomes, while geography has at best weak direct effects*. However, because institutions are the result of a wider variety of factors, both exogenous and endogenous, including human, geography and climate factors, it is more difficult to disentangle causes and effects. For instance, as expressed by Acemoglu et al. (2001), one of the reasons why European settlers did not establish themselves in tropical areas (e.g. sub-Saharan Africa) is the presence of diseases such as yellow fever and malaria which are prevalent in tropical geography/climate conditions.

On the other hand, it is difficult to identify appropriate Institution-related indicators that are truly statistically independent of GDP. Of the six indicators considered by Kaufmann et al. (2009) – *Voice and Accountability*, *Political Stability and Absence of Violence*, *Government Effectiveness*, *Regulatory Quality*, *Rule of Law*, and *Control of Corruption* – their individual linear correlation coefficient with GDP ranges between 0.6 and 0.9, and predominantly at the higher end of this range. For instance, *Rule of law*, as used in Rodrik et al. (2004) has a coefficient of ca. 0.8 with GDP. More importantly, however, a key limitation with institution indicators is that they are not available globally at a resolution finer than country averages, which is essential for the work performed here.

The most attractive feature of trying to explain economic development through geography and climate factors is that these are essentially exogenous elements. In fact, *because weather is exogenous and random in most economic applications, it acts like a "natural experiment"* thus allowing the identification of relationships between economic outcomes and meteorology in a scientific way (Angrist and Krueger 2001, Auffhammer et al. 2013).

In this study we take advantage of a high-resolution global GCP data set (Nordhaus and Chen 2016). The lack of a granular GCP global map has prevented researchers from intimately assessing the connection between geography/climate and economic activity. And although many attempts have been made (Dell et al. 2014 and references therein), studying their link at the level of country-average, i.e. using GDP, is sub-optimal. In fact, climate variables can vary wildly within a country. This is certainly true for very large countries like USA, China or Australia, but even for smaller countries like Italy, with a distinct north-south gradient in both climate and productivity, relying on GDP is not satisfactory (Eckaus 1961).

Another important extension introduced here is the use of a non-linear model. Most of the current literature simulates GCP variations using (multi)linear models, including Mitton (2016). Although practical and reasonable, linear models have major limitations, particularly in a complex non-linear problem like the modelling of economic activity.

The next section, 2, presents the data sets used, while the methodology is discussed in Section 3 and results are examined in Section 4. A summary and discussion are provided in Section 5.

## Data

The underlying meteorological data are provided by the ERA-Interim reanalysis. This reanalysis is produced by the European Centre for Medium-Range Weather Forecasts (ECMWF) and is described in Dee et al. (2011). Here we use data from 1979 to 2016, i.e. most of ERA-Interim. Its main features are a horizontal resolution of 0.75° by 0.75°, and a temporal resolution which varies between 3 hours and 6 hours, depending on the variable (see Table 1). Also, several derivatives of the meteorological variables considered, and listed in Table 1, are used.

The globally gridded GCP data used here comes from the Global Gridded Geographically Based Economic Data (G-Econ, https://gecon.yale.edu/data-and-documentation-g-econ-project), Version 4 (Nordhaus and Chen 2016); see also an early version, for 1990 only, in Nordhaus (2006). This dataset contains derived one-degree grid cells of GCP data for the years 1990, 1995, 2000 and 2005. To allow for an easier direct comparison across the globe, we consider GCP at Purchasing Power Parity and per capita. Per capita GCP (hereafter referred to as GCP-PC) is computed by dividing GCP by the population in each cell, also provided with the G-Econ dataset.

While the main use of the G-Econ is the gridded GCP data, this dataset also includes geographical parameters such as area of grid cells, distance to coast, elevation, vegetation and soil types, distance to rivers, which are also used in our analysis, along with the climate variables (see Table 2). All geography parameters are fixed for the period covered by this study – possible local variations particularly in vegetation or soil types, e.g. around urban centres, do not appear to be critical (see later). To ensure geographical overlap with G-Econ, ERA-Interim is retrieved at 1°, namely at a slightly lower resolution than its original 0.75°.

## Assumptions and Methodology

The main focus of this work is the investigation of the casual relationship between the main features of meteorological variables over a few decades and the corresponding GCP, rather than the (concurrent) correlation between meteorological variables and GCP. These two objectives require different approaches. Specifically, in the first case statistical properties such as seasonal variations of meteorological variables are used. It is this type of features, namely the changes in variables, that we want to analyse in addition to the more standard statistics such as the mean (of e.g. temperature). Accordingly, the main assumption here is that the statistics of both the meteorology and GCP are stationary over the period considered, namely 1979-2016 for the meteorological variables, and the four years of G-Econ, 1990, 1995, 2000 and 2005 for the GCP. Clearly within any given period the climate will vary, but the wide-ranging statistical characteristics we use here take into accounts such variations. More importantly, GCP changes in parts of the globe, particularly for China, over the G-Econ period. Accordingly, the stationarity assumption for GCP has been tested by substituting it with GCP plus and minus one standard deviation (computed using the four years available), respectively: in both cases, the main predictors are the same as for the (mean) GCP, thus confirming the validity of this assumption (not shown).

**Table 1.** Meteorological variables, used in this study, as derived from the ERA-Interim reanalysis. The following statistics have been computed for all variables: Mean, 1st Quartile (bottom Q), Median, 3rd Quartile (top Q), Standard Deviation of the original time series (SD), Standard Deviation of monthly means (SD S, representing intra-annual variations) and temporal variations (representing short-term 'gradients'). The latter are computed according to daily or 6-hourly steps (see column Step). For 6-hourly variables, increments are increased by 10% at each subsequent steps (out to 5 steps, i.e. 30 hours) and by 15% for daily steps (out to 5 days). For air temperature (at 2 m height), also daily excursions are calculated using two additional variables, $T_{min}$ and $T_{max}$, available at 6-hour intervals.

| Climate Variable | Unit | 'gradient' | Step |
|---|---|---|---|
| Mean Sea Level Pressure (MSLP) | hPa | 750 | 6-hr |
| Wind Speed at 10m (UV10) | m s$^{-1}$ | 2.5 | 6-hr |
| Air Temperature (T2) | °C | 5 | Daily |
| Air Temperature daily excursion (DT) | °C | 2.5 | Daily |
| Dew Point Temperature (D2) | °C | 7 | Daily |
| Precipitation (TP) | mm | 5 | Daily |
| Relative Humidity (RH) | % | 20 | Daily |
| Solar Radiation (SR) | W m$^{-2}$ | 100 | Daily |
| Sunshine Duration (SUND) | hr | 3 | Daily |

**Table 2.** Geography parameters. Aside from latitude and elevation (the latter has been taken from the ERA-Interim reanalysis), the other variables are from the G-Econ dataset.

| Geography Variable | Unit | Geography Variable | Unit |
|---|---|---|---|
| Latitude | Degrees | Distance to Major River | km |
| Elevation | m | Distance to River | km |
| Distance to coast 1 | km | Distance to Ocean | km |
| Distance to coast 2 | km | Vegetation category | category [0-31] |
| Distance to Lake | km | Soil category | category [0-250] |

Because a single target (or predictand), namely GCP for each geographical location, is required, this is taken as the average of the four years in G-Econ. To reduce the noise in the analysis, grid cells where the GCP is smaller than 1 USD are not considered. Also, as normally done (e.g. Rodrik et al. 2004), GCP is converted into a logarithm (base 10). The (log of the) GCP thus computed is shown in S1 Figure while the corresponding GCP-PC can be seen in Figure 1. Its Probability Density Function (PDF) is shown in Figure 2.

As already remarked, the focus of this work is the assessment of the dependency of GCP-PC on climate and geography variables. Formally this can be expressed as:

$$\text{GCP-PC} = f(\text{Geography, Climate}) + \underbrace{h(\text{Institutions, National Resources, ...}) + \varepsilon}_{\text{GCP-PC} = f(\text{Geography, Climate}) + \delta} \quad (1)$$

where $\varepsilon$ is the error term, which captures factors controlling GCP-PC not accounted for by the other terms, therefore including errors in GCP, climate and geography variables; $\delta$ is a term not explicitly modelled in this work. By assessing the spatial geographical patterns of $\delta$ it may be possible to identify the source of potential mismatches between GCP-PC and the geography and climate predictors used here through the modelled function, *f*.

The main tool used in our analysis is a non-linear statistical model called Random Forests (RF). This is a well-known and popular method consisting of a set of decision trees built to minimise their correlation (Breiman 2001, Hastie et al. 2008). The model has been chosen after a comparison with other models, Gradient Boosting (GB, a non-linear statistical model, Hastie et al. 2008) and the multi-linear (ML) regression model. The RF model also provides the importance of predictor variables, a very useful and robust diagnostic. Moreover, the issue of overfitting due to collinearity of variables (or predictors) is considerably reduced, or even eliminated, with the RF (Breiman 2001). The justification for using a non-linear model stems from the complex relationships between any one variable and GCP PC. These have been assessed through scatter plots, also reflected in the low correlation coefficients (see later Table 4) as well as, critically, by the low performance of the multi-model approach (see Table 3).

We use around 120 geography and climate predictors (see Tables 1 and 2). By construction many of these predictors are highly correlated. However, since the RF model can deal very well with collinearities (Breiman 2001), correlated variables are not eliminated. Instead, by retaining them we allow for multiple (correlated) predictors to emerge when assessing the variable importance ranking. If two or more highly correlated variables turn out to be the most important, this result will signal the relevance of that group of predictors. Indeed, if we were to remove collinearities a priori we would not be able to identify recurrent important variables (this was tested by eliminating highly correlated variables and retaining only one amongst them).

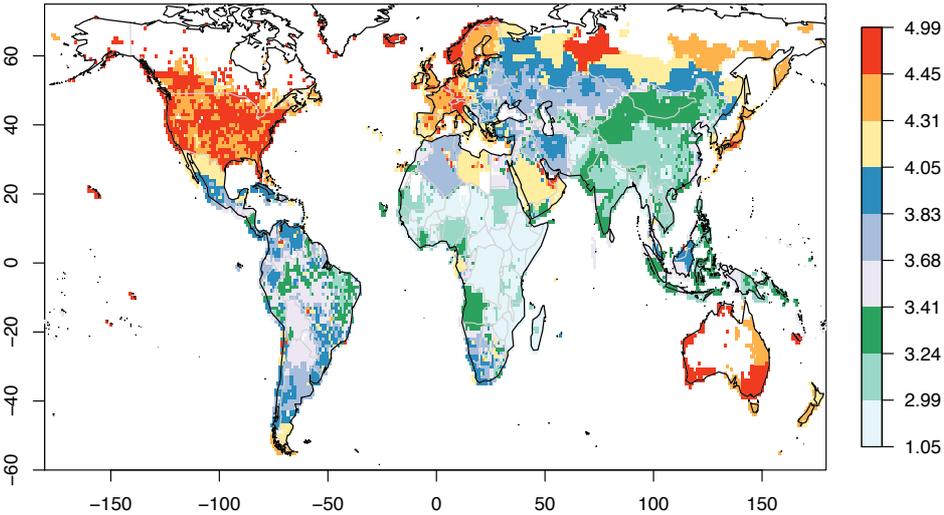

**Figure 1. Map of log Gross Cell Product per capita.** GCP per capita in $\log_{10}$ (k USD/person). Note that the colour palette has been divided into three sets of three colours each, which represent the three terciles, whose thresholds are 3.41 and 4.05, respectively (cf. Figure 2). Source G-Econ (Nordhaus and Chen 2016).

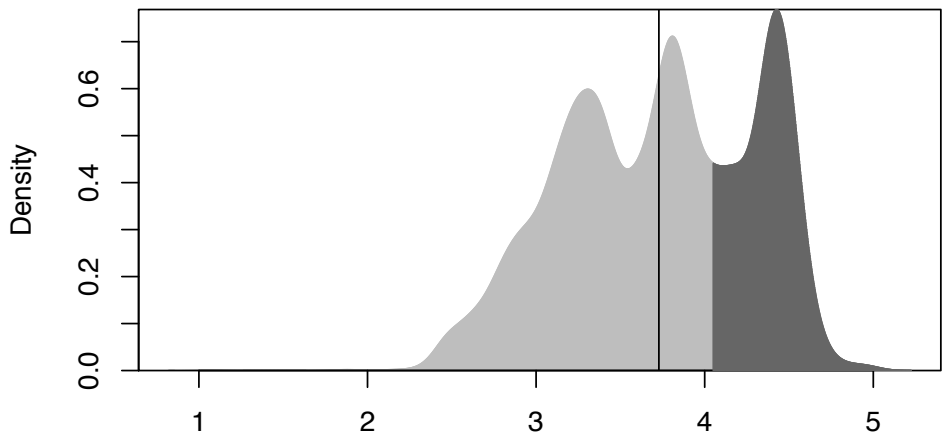

**Figure 2. PDF of Gross Cell Product per capita.** Probability Density Function (PDF) for $\log_{10}$ of GCP-PC. The black vertical line shows the mean of the distribution which is 3.73 (equivalent to about $5350/person). The PDF displays an interesting tri-modal behaviour (cf. also Figure 1): the first mode (between 3.05 and 3.55) corresponds to most of East and South Asia, Central-Northern Brazil, and parts of Western Africa; the second mode (between 3.55 and 4.1) covers Eastern Europe, Northern and Southern Africa, most of Central America and the rest of South America; the third mode (between 4.1 and 4.65) covers North America, Western Europe, parts of Russia, the Arabian peninsula, Japan and coastal Australia. The darker grey denotes the upper tercile.

Predictors are ranked using the RF importance variable feature, according to the following procedure (see also S2 Figure). First we train the RF model with the complete set of all the 121 variables, by running a 1000-member ensemble. The top 10 variables in terms of importance are retained for the next step. A parallel selection, using a sub-sampling approach is adopted. With this approach the RF model is trained with only 20 variables, sub-sampled via a bootstrapping methodology, with an ensemble of 300 realisations. In this case too, the top 10 variables in terms of importance are retained. The two sets of top 10 variables computed with these two separate steps are then pooled together. The final step in the variable selection consists in ranking the two sets, and retaining, the top 10 overall variables, by means of a k-fold (with k equal to 5) cross validation procedure, and minimizing the Mean Absolute Error (MAE). The k-fold is commonly used to test the performance of a prediction model, particularly when there is not independent data for validation (Hastie et al. 2008). Essentially, a sample is divided into k equally populated randomly selected sub-samples. In turn, (k-1) sub-sample(s) is/are used for the training and the remaining sub-sample is used for the prediction. The most typical values of k are 5 or 10.

Four samples are considered:
1. All grid points (namely most of the world);
2. Grid points in the top (or upper) third (or tercile) of the GCP-PC distribution;
3. Grid points in the middle tercile of the GCP-PC distribution;
4. Grid points in the bottom (or lower) tercile of the GCP-PC distribution.

In addition we also use the predictive power of the RF model to reconstruct the GCP-PC global map given the retained top geography and climate predictors, utilising the *k*-fold approach.

## Variable importance and GCP Per Capita prediction

### Global GCP Per Capita

We first compare the performance of the three statistical models (RF, GB and ML). Specifically, we run a comparison between RF and GB keeping all the 121 variables, using the k-fold approach, as well as with the top 10 predictors. The ML model is tested only with the 10 selected predictors to avoid collinearities issues (see Table 3). As shown in Table 3, the RF's error is less than half that of GB in terms of the normalised Mean Absolute Error (nMAE, i.e. MAE divided by the standard deviation of the global log10 GCP-PC distribution). Similarly to the case with all variables, the RF performance for the top 10 predictors is more than twice as better than that of GB. Note however that the first comparison, with all variables, is more robust as the selection of 10 predictors has been made based on the RF model. ML performs considerably worse than either RF and GB: the nMAE for ML is almost four times larger than that of RF. As an additional check, the performance of the RF model is also tested using the out-of-bag (OOB) prediction statistics (this is computed using withheld data within the sample used by the RF trees, and a useful benchmark).

The ranked top 10 variables are shown in Table 4 and in Figure 3. The latter shows how the nMAE, and the correlation, levels off after six-seven predictors, even if some small error reductions are seen with the subsequent predictors. Our analysis confirms that *latitude* is the dominant predictor for GCP-PC. However, it is also important to note that the nMAE with latitude only is around 0.65, hence relatively large, compared to the asymptotic value of ca. 0.15, obtained when at least six predictors are used. The limitations of using latitude-only are also apparent from the geographical reconstruction map (see later Figure 4, top left). Somewhat unexpectedly, however, the second most important variable is a meteorological variable, and specifically MSLP SD S, namely the month-to-month variations of mean sea level pressure (or just air pressure). This variable reduces the nMAE to 0.44, when added to latitude. Not only is this meteorological variable more important than other geography variables, critically it is also not one that is normally considered in the literature (typically mean temperature or precipitation). And while air pressure technically encapsulates information about temperature, it is influenced by several other meteorological variables, like humidity. Importantly, this is not the standard *average* variable but a measure of variability, in this case an indication of seasonal (or intra-annual) variations. Also, while it would be inaccurate to identify MSLP SD S directly with specific meteorological phenomena such as the passage of a storm, this statistical measure contains the signature of storms. Therefore, this result seems to corroborate the more geographically and temporally limited finding of Huntington.

The third most important variable is a geography variable, the distance from a major river. This variable reduces the nMAE to 0.28. The implications of the proximity of a major river are evident from a transport, recreational, etc. view point, but it is also worth mentioning that a link exists between rivers and environmental conditions such as water availability or soil conditions, therefore with clear connections to meteorological variables such as precipitation, temperature and humidity. The fourth variable is another meteorological variable, Dew-Point SD S, namely the month-to-month variations of dew point temperature. This variable reduces the nMAE to ca. 0.23.

**Table 3.** Comparison statistics for different number of variables and different models (RF, GB and ML) and for all grid points (first two columns), the top tercile of GCP-PC (third and fourth columns, the middle tercile of GCP-PC (fifth and sixth columns), the bottom tercile of GCP-PC (last two columns). OOB stands for Out-of-bag prediction (a feature of the RF model). The top 10 predictors have been selected using the procedure described in the Methodology section. The row with the top 10 predictors for RF, which is used in the rest of the study, has been highlighted in grey.

|  | All grid points | | Top tercile | | Middle tercile | | Bottom tercile | |
|---|---|---|---|---|---|---|---|---|
|  | nMAE | CORR | nMAE | CORR | nMAE | CORR | nMAE | CORR |
| RF 121 (all) variables (OOB) | 0.152 | 0.970 | 0.276 | 0.893 | 0.386 | 0.840 | 0.251 | 0.909 |
| RF 121 (all) variables (5-fold) | 0.162 | 0.966 | 0.291 | 0.885 | 0.403 | 0.830 | 0.262 | 0.906 |
| GB 121 (all) variables (5-fold) | 0.346 | 0.874 | 0.449 | 0.755 | 0.526 | 0.699 | 0.386 | 0.822 |
| RF top 10 predictors (5-fold) | 0.155 | 0.970 | 0.268 | 0.894 | 0.383 | 0.842 | 0.263 | 0.902 |
| GB top 10 predictors (5-fold) | 0.368 | 0.862 | 0.500 | 0.695 | 0.534 | 0.690 | 0.417 | 0.800 |
| ML top 10 predictors (5-fold) | 0.600 | 0.679 | 0.764 | 0.314 | 0.730 | 0.453 | 0.661 | 0.533 |

**Table 4.** Top 10 predictors and (*) their one-to-one linear correlation coefficient with GCP-PC for the cases of all-grid-point, top tercile, middle tercile, and bottom tercile. Also, as a reference, the correlation between latitude and MSLP SD (namely, the meteorological variable, of the top 10, with the highest correlation with GCP-PC, and also with latitude) for the all-grid point case is 0.70. Green cells indicate predictors in common with the four cases, light blue in common with three cases and purple in common with two cases. Highlighted in salmon are the largest one-to-one correlation values for each of the four cases.

|  | All grid points | | Top tercile | | Middle tercile | | Bottom tercile | |
|---|---|---|---|---|---|---|---|---|
| Rank | Predictor | Corr.* | Predictor | Corr.* | Predictor | Corr.* | Predictor | Corr.* |
| 1 | Latitude | 0.35 | MSLP top Q | 0.05 | Latitude | 0.24 | Latitude | 0.30 |
| 2 | MSLP SD S | -0.05 | Dist. M. River | 0.06 | MSLP SD S | 0.08 | Dist. M. River | -0.13 |
| 3 | Dist. M. River | -0.07 | Latitude | 0.06 | Dist. M. River | -0.09 | Dist. Ocean | -0.03 |
| 4 | Dew Point SD S | 0.23 | Dew Point SD S | 0.00 | Dew Point SD S | 0.31 | MSLP SD S | 0.32 |
| 5 | Dist. River | 0.06 | MSLP SD | 0.07 | MSLP SD | 0.35 | Dist. Lake | 0.19 |
| 6 | Dist. Ocean | -0.08 | Dist. River | -0.03 | Dist. Lake | -0.27 | Dist. River | 0.17 |
| 7 | Dist. Lake | 0.00 | MSLP SD S | -0.22 | MSLP top Q | 0.29 | Precip. SD S | -0.18 |
| 8 | MSLP SD | 0.46 | Dist. Lake | -0.05 | Dist. River | -0.16 | Dew Point SD S | 0.20 |
| 9 | MSLP bottom Q | 0.00 | Dist. Ocean | -0.06 | Precip. SD S | -0.25 | Dist. Coast | -0.07 |
| 10 | Precip. SD | -0.16 | Solar Rad -ve (1) | 0.18 | Dist. Coast | 0.12 | Wind speed SD S | -0.04 |

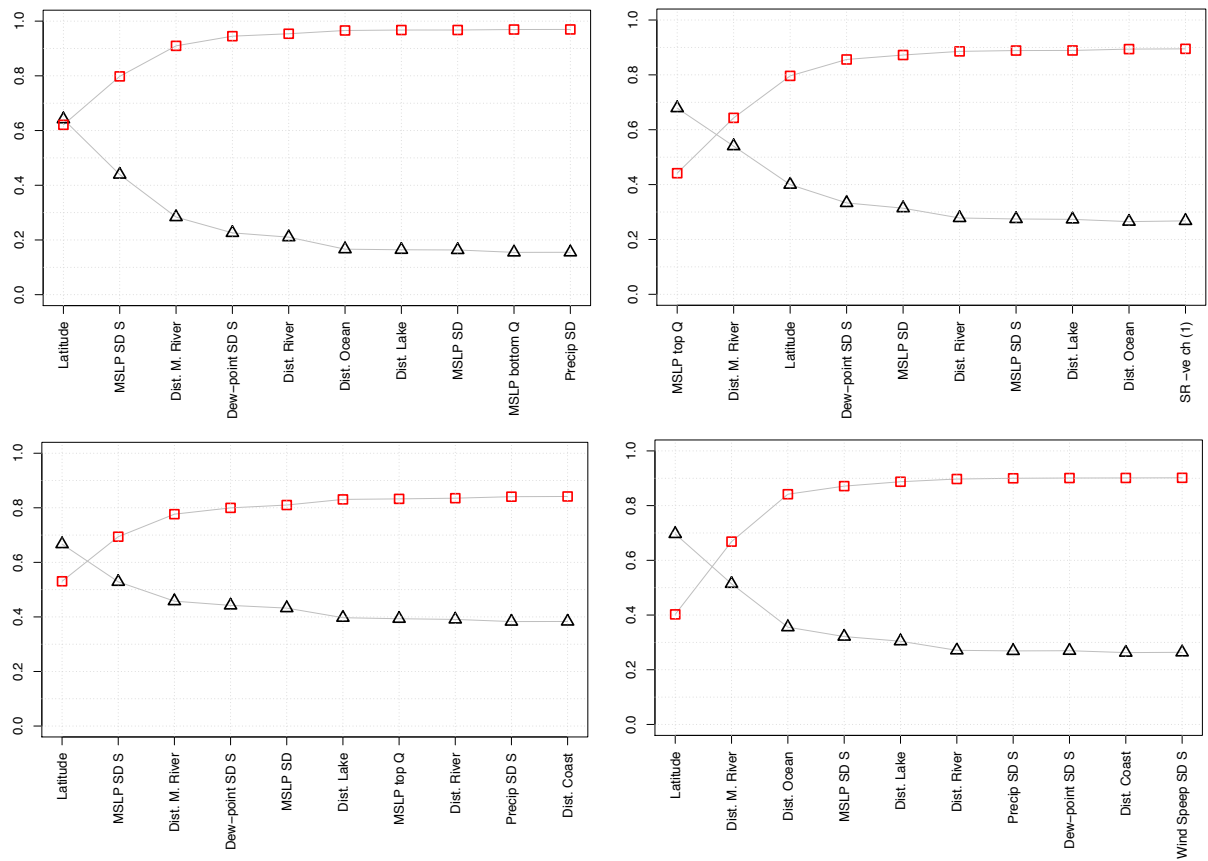

**Figure 3. Summary statistics of GCP-PC prediction.** Statistics of $\log_{10}$ GCP-PC prediction using 5-fold cross validation, adding one predictor at the time in the order given and up to 10 predictors. For instance, in the top left panel 'MSLP SD S' means that 'MSLP SD S' and 'Latitude' are used together (cf. also Table 4). Top left: all grid points. Top right: top tercile of GCP-PC. Bottom left: middle tercile of GCP-PC. Bottom right: bottom tercile of GCP-PC. Metrics used are: linear correlation (red squares) and normalised Mean Absolute Error (nMAE, black triangles). nMAE is equal to 0.56 for all grid points, 0.17 for the top and mid terciles, 0.26 for the bottom tercile.

Despite having temperature in its name, dew point temperature, rather confusingly, is not a temperature. While it also depends on temperature, it is closely associated with relative humidity. Similarly to the case of air pressure, we see here again that the 'traditional' mean temperature and precipitation, while indirectly influencing Dew-Point SD S are not selected by the RF model as the most important variables. Instead, *variations*, rather than mean values, appears to be emerging as most critical factors for explaining the geographical distribution of GCP-PC.

As with the third variable, the fifth (distance to a river), sixth (distance to an ocean) and seventh (distance to a lake) most important variables are of geographical nature, as opposed to (purely) meteorological. Together these three variables further reduce the nMAE to just over 0.16. The last three variables, in the top ten list, are all of meteorological nature and in two cases again related to variations – the overall variations of air pressure (the eighth variable) and of precipitation (the tenth variable). The ninth variable is somewhat different and relates to the distribution of air pressure, namely the bottom quartile of the pressure distribution at any given location. The nMAE is reduced to 0.155 with the tenth variable, which are however minimal adjustments to the nMAE of 0.16 obtained with the top six variables. This means that for the purposes of explaining and reconstructing the all-grid points GCP-PC, six predictors seem to be sufficient. Note also that air pressure enters the top ten list in three different ways: month-to-month variations, overall variations and bottom quartile of its distribution. This is a clear indication that air pressure is a critical variable for explaining the GCP-PC.

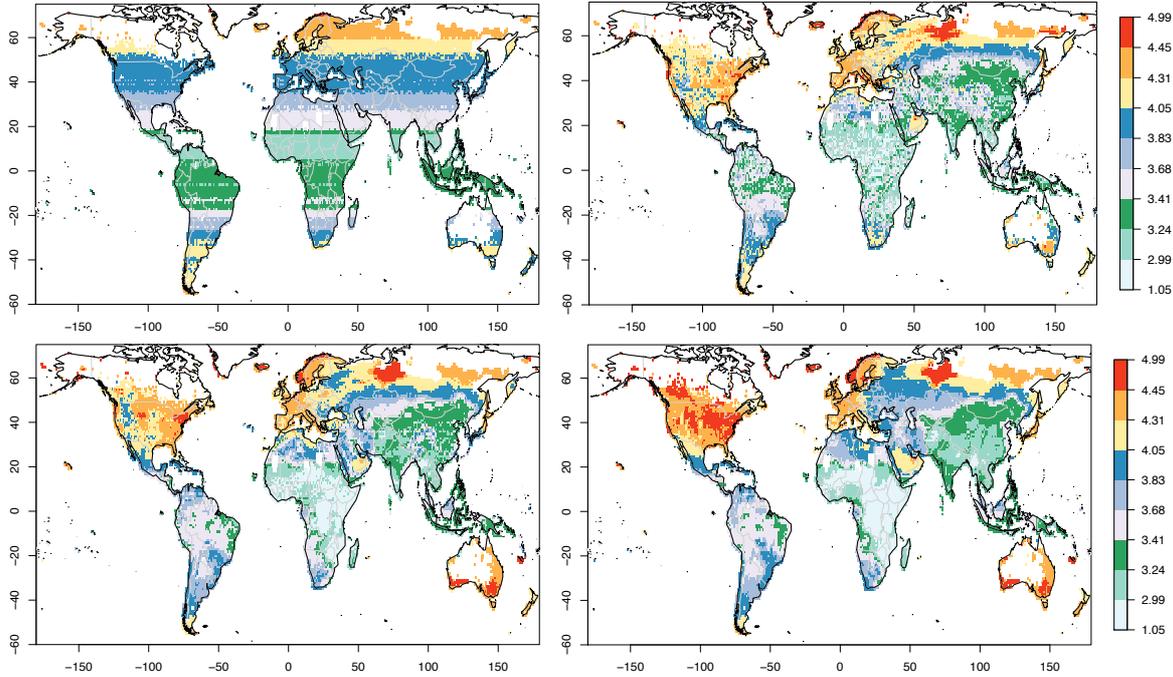

**Figure 4. Maps of GCP-PC prediction.** GCP-PC prediction ($\log_{10}$ of k USD/person) using the top predictors (see Table 4). Top left: top predictor. Top right: top two predictors. Bottom left: top three predictors. Bottom right: top six predictors. Compare with the actual GCP-PC in Figure 1.

The maps with the geographical reconstruction of GCP-PC are shown in Figure 4. The top left GCP-PC map only uses latitude (the top variable) as its predictor. This is apparent from the zonal stripes, and with GCP-PC generally increasing with the absolute value of latitude. It is also evident that using latitude by itself, it is not possible to capture regional variations. These are introduced when the second predictor (MSLP SD S) is also considered. Now the regional fit is considerably improved, to the extent that areas in the Arabian peninsula, but also India, Central Africa and several other parts of the globe, now display a higher GCP-PC than other locations at the same latitude, and captured by the latitude predictor alone. This improvement is more clearly shown in Figure 5 (top left). Here the change in prediction from one step to the next is expressed as the difference between the individual absolute departures from the actual GCP-PC values for prediction 'n-1' minus prediction 'n' and calculated as follows:

$\Delta$ GCP-PC = | Prediction (n-1) – GCP-PC | – | Prediction (n) – GCP-PC |     (2)

A positive (negative) value means that the additional predictor at step 'n' is improving (deteriorating) the prediction obtained at the previous step 'n-1'.

Although the introduction of the second predictor generally improves the fit, there are also areas where the fit deteriorates, such as in Eastern Europe, North West Africa, Southern Africa, South Western USA, and North Eastern Australia. Some of these, particularly the latter one, are rectified when the third predictor (distance to a major river) is used (bottom left in Figure 4 and top right in Figure 5). Others, specifically Eastern Europe, North West Africa and South Western USA, then improve with the fourth predictor (Dew Point SD S) as seen in the bottom left panel of Figure 5. The bottom right panel in Figure 4 shows the geographical fit with the top six predictors, after which the error decreases only marginally (cf. Figure 3). This is confirmed by the bottom right panel of Figure 5 which shows that the change in GCP-PC prediction when the top seven variables are used (compared to the top six) is close to zero in most areas of the globe.

While some noticeable differences between the predicted (using the top 10 predictors) and the measured GCP-PC can be seen (Figure 6 top left), these appear to be due, for the most part, to non-systematic errors. Possible exceptions, with consistent discrepancies either over a single country or a relatively large area, are: Gabon (central West Africa), South-West USA (parts of California and Arizona), and a few Balkan countries (Albania, Montenegro, and Bosnia and Herzegovina). Specifically, the latter area displays a negative bias, with the geography-climate prediction underestimating GCP-PC, and vice-versa in the former two cases. According to Equation (1) these discrepancies may indicate that the combined contribution of institutions and national resources (term $\delta$) is not negligible: this term would be positive in the case of the Balkan countries above, with institutions and/or national resources therefore boosting economic productivity compared, and negative in the other two areas. However, as we will show later, these discrepancies are not present when smaller sections of the GCP-PC data (i.e. split into terciles) are modelled, which may imply that the results for the all-grid case exhibit a minor shortcoming in the RF statistical modelling with a very large amount of data being fit (nearly 13,000 grid points) and/or in the GCP-PC data, particularly for countries where the reporting of economic productivity figures is less established.

### Top tercile of GCP Per Capita

Here we subsample the GCP dataset by retaining only the top tercile of GCP-PC. The RF error again levels off after six or seven variables. From an initial nMAE of 0.68 with one predictor, this reaches a value of 0.28 after six variables and only improves by about 0.01 with the addition of the following four predictors – the nMAE with the top ten predictors is 0.27 (Figure 3 and Table 3). Two main features stand out in the case of the top tercile, compared to the all-grid-point case. The first is that latitude is not the most important predictor; rather it now ranks third, after the top quartile of air pressure (MSLP top Q) and the distance from major rivers. The second feature is that meteorological variables have overall increased in their importance, with three variables in the top six compared to two in the all-location case (Figure 3). The top predictor, MSLP top Q, indicates that high air pressure is a key factor in determining the GCP-PC for countries in the top tercile. Also, and as in the case of all-grid-point, variations in climatic variables, rather than their mean, are still very important: the month-to-month variations for the dew-point temperature (essentially relative humidity) and the overall variations of air pressure (MSLP SD) rank in the top five most important variables. The nMAE is reduced to 0.31 after the fifth variable, MSLP SD. Next, in sixth position, distance to rivers is again an important predictor, with a further reduction of nMAE by ca. 0.04. It is also interesting to note the presence of an unusual variable in the top ten predictors: this is the (negative) change (or gradient) in solar radiation from one day to the next. Even though this variable does not incrementally contribute much to the MAE reduction, after the previous nine predictors, the fact that day-to-day variations (in this case reductions) in solar radiation appears to influence GCP-PC may be worth of further investigation.

Table 4 directly compares the top ten predictors in both the all-grid-point and top tercile cases. It is evident the large overlap amongst them, with eight variables in common (grey boxes), and a ninth variable being of a similar nature, MSLP top Q and MSLP bottom Q. Only the tenth variables in either case are different, though in both cases of meteorological nature. While the overlap is strong evidence of the robustness of the methodology adopted here, there is an interesting change in ranking, from the all-grid-point to the top tercile, with latitude decreasing in importance while meteorology enhancing its role, as already remarked above.

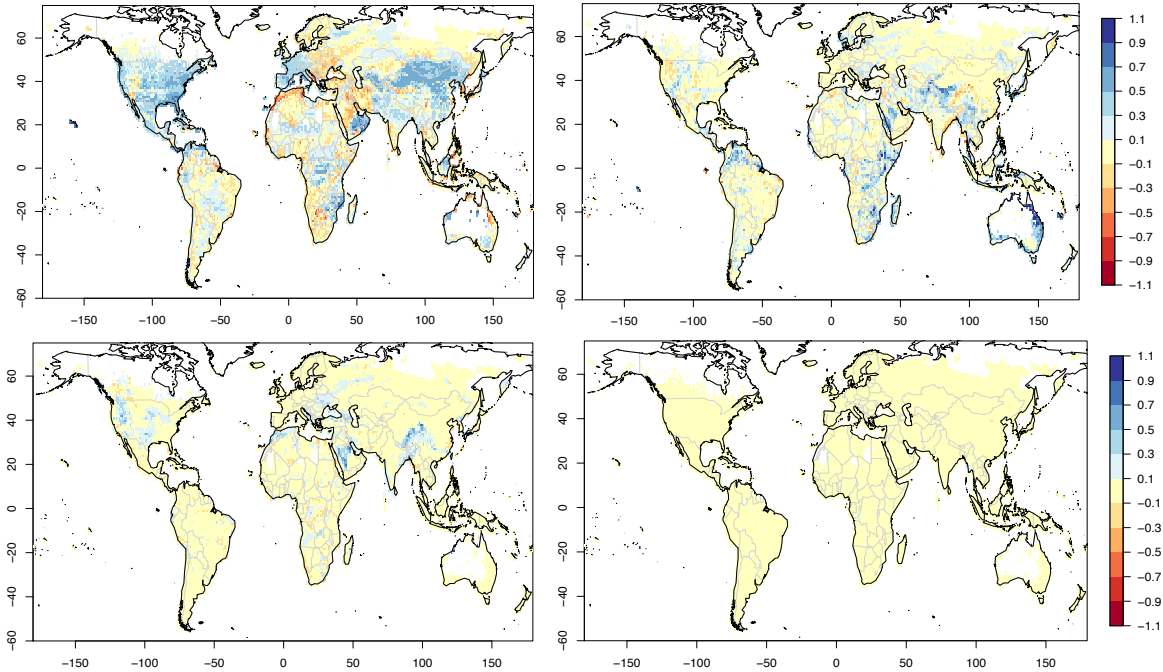

**Figure 5. Maps of GCP-PC prediction departures.** Changes in prediction expressed as the difference between the individual absolute departures from the actual GCP-PC values for step 'n-1' minus step 'n'. A positive (negative) value means that the additional predictor is improving (deteriorating) the prediction. Top left: Change from step 1 to 2. Top right: Change from step 2 to 3. Bottom left: Change from step 3 to 4. Bottom right: Change from step 6 to 7.

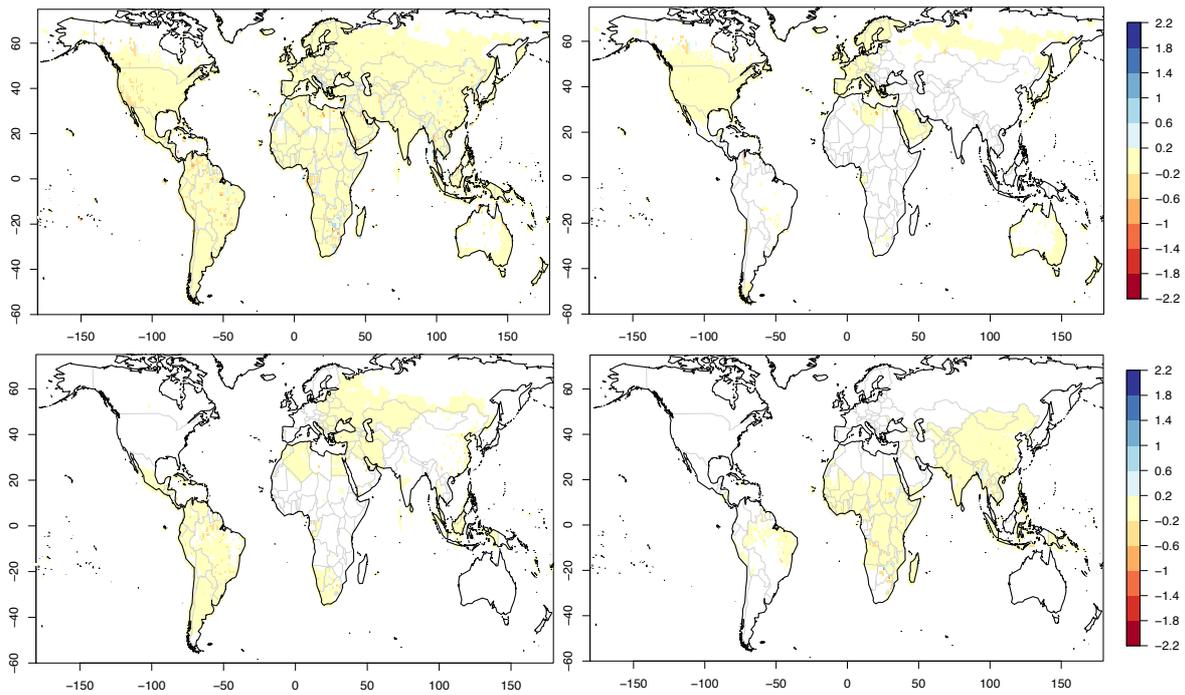

**Figure 6. Difference between predicted and observed GCP-PC.** Difference between predicted (using the top 10 predictors) and observed GCP-PC, namely ($f$ – GCP-PC) in Equation (1). Top left: all grid points. Top right: top tercile of GCP-PC. Bottom left: middle tercile of GCP-PC. Bottom right: bottom tercile of GCP-PC.

The RF model yields again a remarkable geographical prediction/reconstruction of the top tercile of GCP-PC (S3 Figure). While the top quartile of air pressure (top predictor) sufficiently reproduces the overall geographical pattern of GCP-PC, subsequent predictors clearly enhance the fit (S4 Figure). For instance, evident improvements over a few areas in Russia, Libya, the Arabian peninsula and northern/western Australia are seen with the inclusion of the distance from major rivers. Latitude and the month-to-month variations of dew-point temperature further noticeably improve the fit to the GCP-PC, particularly over Libya, southern USA and northern Mexico in the case of latitude, and Arabian peninsula and Spain in the case of dew-point temperature. Subsequent predictors only modify marginally the fit. When compared with the measured GCP-PC, there do not seem to be systematic biases in the predicted (using the top 10 predictors) GCP-PC (Figure 6 top right).

### Mid and bottom terciles of GCP Per Capita

To complement the analysis of the top tercile, we concisely discuss the results for the middle and bottom terciles of GCP-PC. In terms of overall statistics (Table 3), the middle terciles yields slightly worse results than the top tercile across most of the metrics, notably for the RF performance, which is however still considerably superior to GB, and ML. The bottom tercile, instead, yields very marginally better results than the top tercile. Further, in terms of actual predictors, Table 4 shows that there is a high level of agreement for all four cases, with six predictors out of ten in common when all four cases are taken together. When cases are taken in pairs the number of common predictors increases to eight or more predictors. Further, even in the case of the middle and bottom terciles the RF model is capable of reproducing well the observed respective GCP-PC using the top six predictors (S5-S8 Figures). This assessment demonstrates that the statistical, and physical, link between GCP-PC and geography and climate variables is robust. As with the top tercile, there do not seem to be systematic biases in the predicted (using the top 10 predictors) GCP-PC (Figure 6 bottom panels), except perhaps for a couple of areas in the case of the bottom tercile: north-west Myanmar and around the border of Zimbabwe (Figure 6 bottom right), though these are likely due to GCP-PC data issues.

### Patterns of top climate variables

We now qualitatively assess the regional patterns of the top climatic predictors, as a way to gain a better understanding of how such predictors affect GCP-PC. Table 4 shows the linear correlation coefficient between the top ten predictors and the corresponding GCP-PC, for all-grid-point and the three terciles, respectively. It is interesting to note that individually none of the predictors has a very high linear correlation with GCP-PC, in any of the four cases. The highest correlation coefficients are with MSLP SD (0.46), and, expectedly, latitude (0.35). Both of them occur in the all-grid-point cases. The highest correlation in the top tercile case is with MSLP SD S (-0.22), followed, interestingly, by the change in solar radiation (tenth predictor, with 0.18). It is also worth noting the cross-correlation between the predictors with the largest correlation with GCP-PC, namely latitude, MSL SD and MSL SD S. The correlation between latitude and MSLP SD is 0.70; between latitude and MSLP SD S is 0.38; and between MSLP SD and MSLP SD S is 0.60.

The relatively low correlations with GCP-PC are also confirmed when looking at the regional geographical patterns. While there is no marked correlation between these fields and GCP-PC, except perhaps with MSL SD in the all-grid-point case (see Table 4), broadly speaking high GCP-PC is accompanied by values of: MSL SD S between 1 and 4 hPa, MSL SD between 5 and 13 hPa, MSL top Q between 1018 and 1026 hPa, D2 SD S between 4 and 10 °C (Figure 7).

We conclude that while overall a statistical model like RF is able to reconstruct the GCP-PC when the top 6-7 predictors are used, it is more difficult to pinpoint specific predictor's regional features that are responsible for the local GCP-PC, except perhaps with the variations of air pressure (MSL SD).

### Conclusions and Discussion

This work has investigated the role that climate and geography variables have on worldwide economic activity as measured by the per capita Gross Cell Product (GCP-PC), at a fine, 1° by 1°, geographical resolution. We considered two main cases: all global grid points and upper tercile of GCP-PC. We find that eight out of the top ten predictors are in common in these two cases. However, two interesting distinctions are worth highlighting. Firstly, latitude is the top predictor in the first case, but it is less important in the second case (it ranks third). Secondly, climate variables increase in importance, compared to geography variables, in the second case. Specifically, we have seen that month-to-month variations of meteorological variables, as well as their daily variations – particularly mean sea level pressure and dew point temperature – are the main climate predictors that explain economic activity worldwide. Interestingly less than ten variables, and usually six-seven variables explain around 80% of the variance in GCP-PC.

For completeness, the middle and bottom terciles of GCP-PC have also been modelled. The simulations and predictions for these two cases provide further evidence that most of the economic activity, as represented by the GCP-PC, can be explained through a (limited number of) geography and climate predictors, even in areas where the role of Institutions and/or natural resources could be pronounced, or where the GCP-PC data may have shortcomings. Further, and crucially, the results for the three terciles, as well as for the whole distribution, indicate a close agreement amongst each other in terms of the most important explanatory geography and climate predictors, with six variables, out of ten, in common when all four cases together are considered.

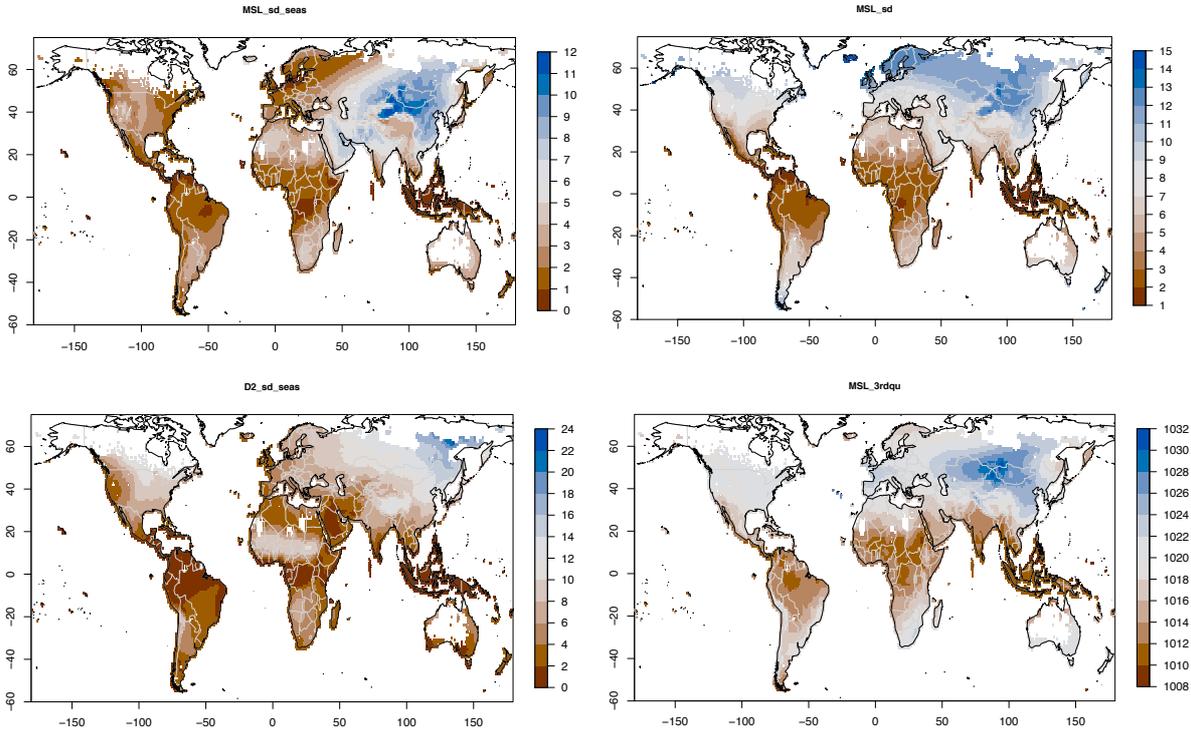

**Figure 7. Maps of main climatic predictors.** Geographical features of the four main climatic (meteorological) predictors: MSL SD S (top left), MSL SD (top right), Dew Point SD S (bottom left), and MSLP top Q (bottom right). There is no marked correlation between these fields and GCP-PC, with the highest correlation being with MSL SD (see also Table 4), but broadly speaking high GCP-PC is accompanied by values of: MSL SD S between 1 and 4 hPa, MSL SD between 5 and 13 hPa, D2 SD S between 4 and 10 °C and MSL top Q between 1018 and 1026 hPa.

While the role of Institutions (and natural resources) has not (explicitly) been accounted for in this study, such robust result, together with the fact that the global GCP-PC can be reproduced with a high level of accuracy by using a set of geography and meteorology variables, provides strong evidence that these are critical factors in explaining fine location-specific economic activities. Our approach could also be used as a basis to investigate the relationship between Institutions and geography and climate factors, if Institution indicators were available at the same, or comparable, geographical resolution as GCP globally.

Our results may have other important implications such as the fact that the relationship between climate and economic activity in the recent past could provide an indication of what the climate conditions were in the distant past in relation to known economically active regions of the world (e.g. the once prosperous Mesopotamia). Conversely, knowing how the climate is projected to vary in the second half of this (XXI) century can give an indication of the possible future economic activity in various parts of the world.

## Acknowledgements

The author would like to thank computer scientist Dr Matteo de Felice for providing expert guidance on the use of statistical approaches. Expert economics feedback by Prof. Arjan Verschoor, Dr Don Gunasekera and Prof. Shaun Vahey were gratefully received.

## Supplementary Figures

These can be downloaded at:

http://www.wemcouncil.org/Output/Geography_Climate_Economic_Activity_ArXiV_SF_Jan2019.pdf